\documentclass[aps,onecolumn]{revtex4}
\usepackage{graphicx}
\usepackage{epsfig}
\usepackage{epstopdf}
\usepackage{amsfonts}
\usepackage{amssymb}
\usepackage{amsbsy}
\usepackage{amsmath}
\usepackage{mathrsfs}
\usepackage{latexsym}
\usepackage{natbib}
\usepackage{bm}
\usepackage{color}
\usepackage[a4paper, total={6.2in, 9.3in}]{geometry}
\DeclareMathOperator{\sech}{sech}
\DeclareMathOperator{\arcsec}{arcsec}

\usepackage{braket}
\usepackage{slashed}
\usepackage{pgfplots}
\numberwithin{equation}{section}

\usepackage{tikz}
\usetikzlibrary{shapes,arrows,shadows}

\begin{document}

\title{Solitons in a class of interacting scalar field theories without $SO(2)$ invariance}

\author{Susobhan Mandal}
\email{sm17rs045@iiserkol.ac.in}

\affiliation{ Department of Physical Sciences, 
Indian Institute of Science Education and Research Kolkata,
Mohanpur - 741 246, WB, India }


\begin{abstract}
In this article, we study kink soliton configurations in interacting scalar field theories containing two fields without $SO(2)$ invariance. We study a class of such theories, the well-known Montonen-Sarker-Trullinger-Bishop model is one of them. These models are interesting since the $U(1)$ current is not conserved in them due to the presence of explicit symmetry breaking terms in the action. The existence of kink soliton configurations is shown in terms of a system of first-order ordinary differential equations. Although $U(1)$ current in these models are non-conserved, our approach is general enough to study soliton configurations in a generic two interacting scalar field theory. We also discuss other benefits of this approach.
\end{abstract}

\maketitle

\section{Introduction}
Soliton configurations in the classical field theories are well-known in the literature of dynamical systems and field theories. The existence of soliton configurations has been discovered in a wide range of physical phenomena \cite{kimball1980kinks, remoissenet1978cooperative, chamon2000solitons, semenoff2008domain}. These solutions are important since they are finite-energy classical configurations that can not be obtained perturbatively from a trivial vacuum configuration. The existence of such finite-energy solutions is crucial in understanding the interplay between the topology of space-time and various physical phenomena. Therefore, it is important to enlarge our understanding of solitons since they are useful in the discovery of new physical phenomena \cite{villari2018sine, 2020CmPhy...3...14S} in low-dimensional physical systems. The analytic solutions of kink solitons in the case of single scalar field theories exist because of the integrability condition shown in \cite{vachaspati2006kinks, rajaraman1982solitons}.

In the literature of solitons, there exist a certain class of interacting field theories containing two fields with the potentials such that a system of first-order ordinary differential equations (ODE), governing the dynamics can be obtained easily. These first-order ODEs are essentially obtained through the minimization of energy density (see \cite{gomes2009extended, de2006expanding, bazeia2002topological}). Among them, there is a class of models in which the dynamical solutions of solitons can be obtained analytically since the system of first-order ODEs can be solved analytically in these cases. The generic action, describing this class of system is of the following form
\begin{equation}\label{0.0}
\begin{split}
\mathcal{A} & =\int d^{2}x\Big[\frac{1}{2}\partial_{\mu}\phi_{1}\partial^{\mu}\phi_{1}+\frac{1}{2}\partial_{\mu}\phi_{2}\partial^{\mu}\phi_{2}-V(\phi_{1},\phi_{2})\Big]\\
V(\phi_{1},\phi_{2}) & =\frac{1}{2}\mathcal{W}_{\phi_{1}}^{2}(\phi_{1},\phi_{2})+\frac{1}{2}\mathcal{W}_{\phi_{2}}^{2}(\phi_{1},\phi_{2}),
\end{split}
\end{equation}
where $\mathcal{W}_{\phi_{i}}(\phi_{1},\phi_{2})=\frac{\partial\mathcal{W}}{\partial\phi_{i}}$ and $\mathcal{W}(\phi_{1},\phi_{2})$ is a smooth function in $(\phi_{1},\phi_{2})$. The Euler-Lagrange equations for the static configurations corresponding to the above action are the following
\begin{equation}\label{0.01}
\begin{split}
\frac{d^{2}\phi_{1}}{dx^{2}} & =\mathcal{W}_{\phi_{1}}\frac{\partial^{2}\mathcal{W}}{\partial\phi_{1}^{2}}+\mathcal{W}_{\phi_{2}}\frac{\partial^{2}\mathcal{W}}{\partial\phi_{1}\partial\phi_{2}}\\
\frac{d^{2}\phi_{2}}{dx^{2}} & =\mathcal{W}_{\phi_{1}}\frac{\partial^{2}\mathcal{W}}{\partial\phi_{1}\partial\phi_{2}}+\mathcal{W}_{\phi_{2}}\frac{\partial^{2}\mathcal{W}}{\partial\phi_{2}^{2}}.
\end{split}
\end{equation}
In these models, the expression for the energy density of static configurations is of the following form \cite{bogomol1976stability, prasad1975exact}
\begin{equation}\label{0.1}
\begin{split}
\mathcal{E} & =\frac{1}{2}\left(\frac{d\phi_{1}}{dx}\right)^{2}+\frac{1}{2}\left(\frac{d\phi_{2}}{dx}\right)^{2}+\frac{1}{2}\mathcal{W}_{\phi_{1}}^{2}(\phi_{1},\phi_{2})+\frac{1}{2}\mathcal{W}_{\phi_{2}}^{2}(\phi_{1},\phi_{2})\\
 & =\frac{1}{2}\left(\frac{d\phi_{1}}{dx}\mp\mathcal{W}_{\phi_{1}}(\phi_{1},\phi_{2})\right)^{2}+\frac{1}{2}\left(\frac{d\phi_{2}}{dx}\mp\mathcal{W}_{\phi_{2}}(\phi_{1},\phi_{2})\right)^{2}\pm\frac{d\mathcal{W}}{dx}.
\end{split}
\end{equation}
The minimum energy configurations (BPS states) can be obtained by solving the following two first-order ODEs
\begin{equation}\label{0.2}
\frac{d\phi_{1}}{dx}=\pm\mathcal{W}_{\phi_{1}}(\phi_{1},\phi_{2}), \ \frac{d\phi_{2}}{dx}=\pm\mathcal{W}_{\phi_{2}}(\phi_{1},\phi_{2}),
\end{equation}
provided the energy $\mathcal{W}[\phi_{1}(\infty),\phi_{2}(\infty)]-\mathcal{W}[\phi_{1}(-\infty),\phi_{2}(-\infty)]$ is finite. The positive and negative signs in the above set of ODEs denote the kink and anti-kink soliton configurations respectively. Further, note that the above ODEs lead only to a soliton solution such that the total energy of the system is minimized. This is a very strong condition. In general, the solutions of solitons require only a finite-energy configuration that demands
\begin{equation}
\left(\frac{d\phi_{i}}{dx}\mp\mathcal{W}_{\phi_{i}}(\phi_{1},\phi_{2})\right)_{x\rightarrow\pm\infty}=0, \ \frac{d\mathcal{W}}{dx}\Big|_{x\rightarrow\pm\infty}=0,
\end{equation} 
equivalent to 
\begin{equation}
\frac{d\phi_{i}}{dx}\Big|_{x\rightarrow\pm\infty}=0=\mathcal{W}_{\phi_{i}}(\phi_{1}(x\rightarrow\pm\infty),\phi_{2}(x\rightarrow\pm\infty)),
\end{equation}
which is a weaker condition. Furthermore, the equations in (\ref{0.2}) can be obtained from (\ref{0.1}) which holds only for the potentials of the form, given in (\ref{0.0}). However, the solutions of the first-order equations in (\ref{0.2}) are the solutions of the Euler-Lagrange equations in (\ref{0.01}) coming from the least action principle provided the same signs are chosen in (\ref{0.2}) but not the vice-versa.

In this article, the existence of soliton configurations is shown in a class of interacting field theories with two scalar fields without $SO(2)$ invariance. Through an example, we show the existence of a system of first-order ODEs, determining the complete dynamics of these field theories. MSTB model \cite{montonen1976solitons, sarker1976solitary, subbaswamy1981intriguing} is chosen as a simple example of this class of scalar field theories which is not $SO(2)$ invariant due to the non-zero mass of a scalar field. On the other hand, for a $SO(2)$ invariant field theory, it is shown that the conserved $U(1)$ current must be vanishing for the existence of a finite-energy soliton configuration. This makes those field theories effectively a single scalar field theory in $1+1$-dimension. In the MSTB model, using the non-conserved $U(1)$ current field variable, we obtain three coupled first-order ODEs which can be solved numerically. Further, it is also shown that the dynamics of every physical observable in this field theory can be obtained from the dynamics of a single scalar field with a soliton configuration. This approach can easily be generalized to the other two interacting scalar field theories, shown through a different example. 

The scalar field theories with $SO(2)$ invariance enjoy global $U(1)$ invariance, hence, $U(1)$-current is conserved. Although in low-dimensional field theories, the $U(1)$ symmetry or the $SO(2)$ symmetry can not be spontaneously broken since it is forbidden by the Mermin-Wagner theorem \cite{mermin1966absence, dobrushin1975absence, gelfert2001absence}, however, under certain circumstances, these symmetries can be broken explicitly \cite{zou2016traveling, arutyunov2008superconductivity, haim2016interaction, rice1965superconductivity, tucker1971onset, maruyama2007u, schmiedmayer2018one, keil2015optical}. These examples play an important role in different areas of condensed matter physics. MSTB is one such simple model in which $SO(2)$ symmetry is explicitly broken due to the non-zero mass of one of the fields. It is shown that given the asymptotic $U(1)$ current to be zero as an initial boundary condition, there exists a non-trivial and non-zero spatial configuration of $U(1)$ current with a finite-energy configuration. 

In order to find a soliton configuration in a given classical field theory, the Euler-Lagrange equations must be solved as a well-posed boundary value problem. Since the Euler-Lagrange equations are the second-order differential equations, they can be equivalently expressed as four first-order differential equations with the same boundary conditions in the field theories with two interacting fields. However, in a certain class of field theories like in (\ref{0.0}), we need to solve only two coupled first-order differential equations to obtain the soliton configurations. In order to find soliton configurations in the field theories beyond this class, we find a system of two coupled first-order differential equations using the $U(1)$ current that must be solved as a well-posed boundary value problem. Moreover, these equations are obtained without imposing any strong condition like in (\ref{0.0}, \ref{0.2}). This approach can also be generalized to the field theories containing more fields.

Now onwards, solutions of kink and anti-kink solitons essentially refer to the static kink and anti-kink solitons. The time-dependent soliton configurations can be obtained by applying a Lorentz boost to the static configurations, given by the following transformation
\begin{equation}
(x-x_{0})\rightarrow\gamma(x-x_{0}-vt), \ \gamma=\frac{1}{\sqrt{1-v^{2}}}.
\end{equation}    

\section{$SO(2)$ invariant interacting scalar field theory}
\subsection{Introduction to model}
Here we study a global $SO(2)$ invariant scalar field theory. There are many such examples of field theories, however, we want to generalize the known $\phi^{4}$ kink configuration. Hence, we choose a system described by the following action
\begin{equation}\label{eqn.0.1}
S=\int d^{4}x\Big[\frac{1}{2}\sum_{a=1}^{2}\partial_{\mu}\phi_{a}\partial^{\mu}\phi_{a}-U(\{\phi_{a}\})\Big],
\end{equation}
where $U(\{\phi_{a}\})$ is chosen to be $U(\phi_{1},\phi_{2})=\frac{\lambda}{4}(\phi_{1}^{2}+\phi_{2}^{2}-\eta^{2})^{2}$, shown in Fig.1. The Euler-Lagrange equations of motion are the following
\begin{equation}
\begin{split}
\partial_{\mu}\partial^{\mu}\phi_{1} & =-\lambda\phi_{1}(\phi_{1}^{2}+\phi_{2}^{2}-\eta^{2})\\
\partial_{\mu}\partial^{\mu}\phi_{2} & =-\lambda\phi_{2}(\phi_{1}^{2}+\phi_{2}^{2}-\eta^{2}).
\end{split}
\end{equation}

\begin{figure}
\begin{center}
\includegraphics[height=7cm,width=9cm]{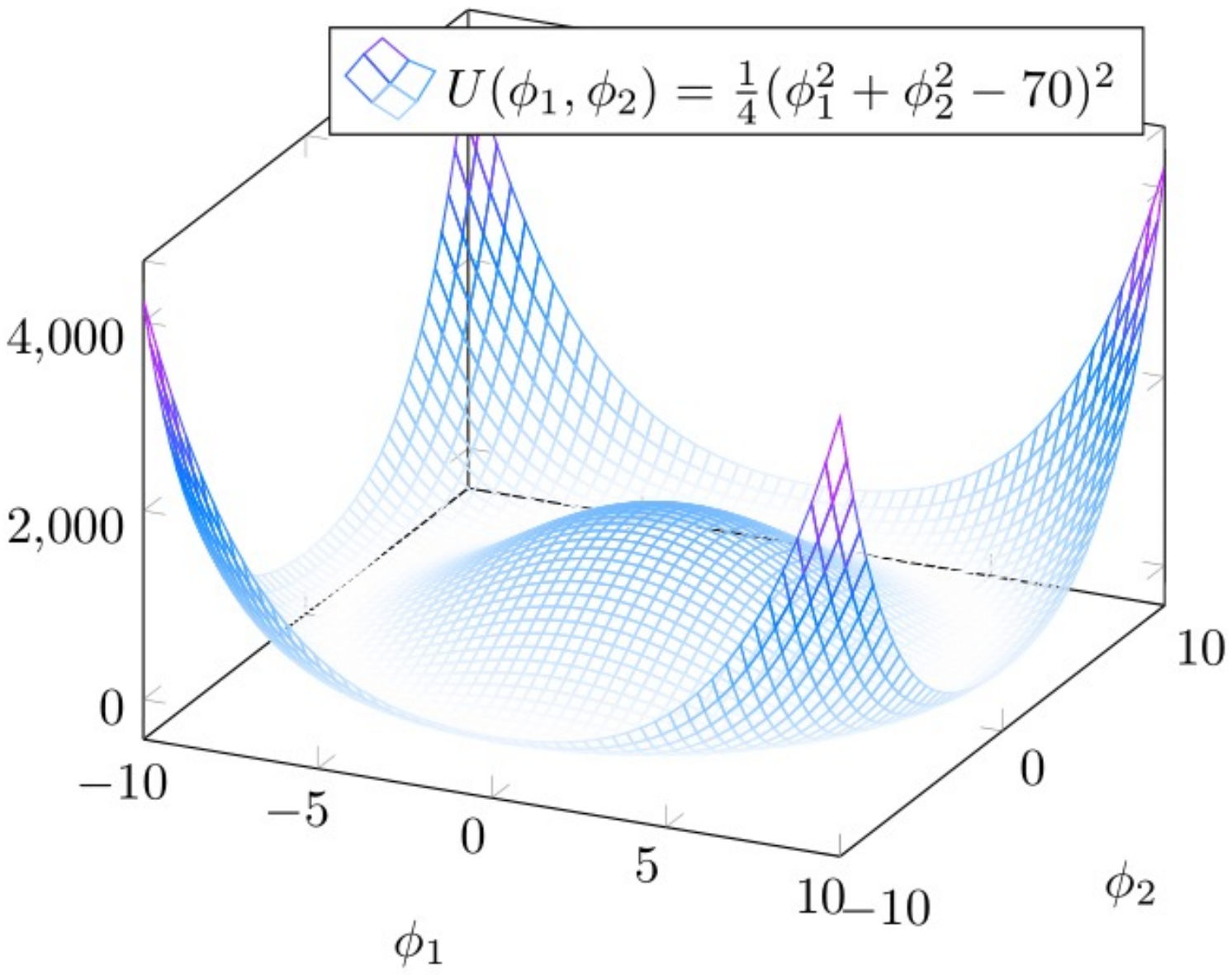}
\end{center}
\caption{Example of a $SO(2)$ invariance potential}
\end{figure}

\subsection{Complex field parametrization}
In this section, the complex field parametrization is introduced in order to show some general results, hold for an arbitrary scalar field theory with two interacting fields. Consider a generic scalar field theory, described by the following action
\begin{equation}\label{action1}
S=\int d^{2}x\Big[\frac{1}{2}\partial_{\mu}\phi_{1}\partial^{\mu}\phi_{1}+\frac{1}{2}\partial_{\mu}\phi_{2}\partial^{\mu}\phi_{2}-U(\phi_{1},\phi_{2})\Big].
\end{equation}
Using the following parametrization
\begin{equation}
\phi=\frac{1}{\sqrt{2}}(\phi_{1}+i\phi_{2}), \ \bar{\phi}=\frac{1}{\sqrt{2}}(\phi_{1}-i\phi_{2}),
\end{equation}
(\ref{action1}) can be expressed as follows
\begin{equation}
S=\int d^{2}x\Big[\partial_{\mu}\bar{\phi}\partial^{\mu}\phi-V(\phi,\bar{\phi})\Big],
\end{equation}
where $V(\phi,\bar{\phi})\equiv U(\phi_{1},\phi_{2})$. The corresponding equations of motion of the above complex field variables are
\begin{equation}\label{EL1}
\begin{split}
\partial_{\mu}\partial^{\mu}\bar{\phi}+\frac{\partial V(\phi,\bar{\phi})}{\partial\phi} & =0, \ \partial_{\mu}\partial^{\mu}\phi+\frac{\partial V(\phi,\bar{\phi})}{\partial\bar{\phi}}=0.
\end{split}
\end{equation}
From (\ref{EL1}), it follows that
\begin{equation}
\begin{split}
\partial_{\mu}\phi\Box\bar{\phi} & +\partial_{\mu}\bar{\phi}\Box\phi+\partial_{\mu}\phi\frac{\partial V(\phi,\bar{\phi})}{\partial\phi}+\partial_{\mu}\bar{\phi}\frac{\partial V(\phi,\bar{\phi})}{\partial\bar{\phi}}=0\\
\implies\partial_{\mu}\phi\Box\bar{\phi} & +\partial_{\mu}\bar{\phi}\Box\phi+\partial_{\mu}V(\phi,\bar{\phi})=0,
\end{split}
\end{equation}
which in static limit becomes
\begin{equation}
\frac{d}{dx}\Big[\partial_{x}\phi\partial_{x}\bar{\phi}-V(\phi,\bar{\phi})\Big]=0.
\end{equation}
Hence,
\begin{equation}\label{5.7}
\partial_{x}\phi\partial_{x}\bar{\phi}-V(\phi,\bar{\phi})=\text{constant}.
\end{equation}
The expression for the energy density of a static configuration is given by
\begin{equation}\label{5.8}
\varepsilon=\partial_{x}\phi\partial_{x}\bar{\phi}+V(\phi,\bar{\phi}).
\end{equation}
The existence of a finite-energy configuration demands that $\partial_{x}\phi|_{x\rightarrow\pm\infty}$ and $\partial_{x}\bar{\phi}|_{x\rightarrow\pm\infty}$ must vanish, and $\phi(x\rightarrow\pm\infty), \ \bar{\phi}(x\rightarrow\pm\infty)$ must become the roots of the potential $V(\phi,\bar{\phi})$. Hence, the equation (\ref{5.7}) becomes
\begin{equation}\label{5.9}
\partial_{x}\phi\partial_{x}\bar{\phi}=V(\phi,\bar{\phi}).
\end{equation}
Further, using (\ref{EL1}), we obtain the following constraint
\begin{equation}
\partial_{\mu}(\bar{\phi}\partial^{\mu}\phi-\partial^{\mu}\bar{\phi}\phi)+\left(\bar{\phi}\frac{\partial V}{\partial\bar{\phi}}-\phi\frac{\partial V}{\partial\phi}\right)=0,
\end{equation}
which shows $j^{\mu}=i(\bar{\phi}\partial^{\mu}\phi-\partial^{\mu}\bar{\phi}\phi)$ is a conserved current if and only if $V(\phi,\bar{\phi})$ is only a function of $|\phi|^{2}=\bar{\phi}\phi$. If $V(\phi,\bar{\phi})$ is only a function of $|\phi|$, then the action (\ref{action1}) is invariant under the action of $U(1)$ group. For a static soliton configuration, the spatial component of current satisfies $j_{x}(x\rightarrow\pm\infty)=-i(\bar{\phi}\partial_{x}\phi-\partial_{x}\bar{\phi}\phi)|_{x\rightarrow\pm\infty}=0$. Furthermore, even if the $V(\phi,\bar{\phi})$ is not $U(1)$ invariant, the following condition still holds.
\begin{equation}
\left(\bar{\phi}\frac{\partial V}{\partial\bar{\phi}}-\phi\frac{\partial V}{\partial\phi}\right)\Bigg|_{x\rightarrow\pm\infty}=0.
\end{equation}
The above quantity in the parenthesis is nothing but the variation of $V(\phi,\bar{\phi})$ \textit{w.r.t} $\alpha$ under the infinitesimal version of the transformation $\phi\rightarrow\phi e^{-i\alpha}, \ \bar{\phi}\rightarrow\bar{\phi}e^{i\alpha}$. This essentially implies that the existence of a soliton configuration demands the potential $V(\phi,\bar{\phi})$ must be $U(1)$ invariant in $x\rightarrow\pm\infty$. This is consistent with the example of potential, shown in (\ref{eqn.0.1}).

Using the polar parametrization $\phi_{1}(x)+i\phi_{2}(x)\equiv\rho(x)e^{i\theta(x)}$ and $\phi_{1}(x)-i\phi_{2}(x)\equiv\rho(x)e^{-i\theta(x)}$, equation (\ref{5.9}) can be expressed as
\begin{equation}\label{5.13}
\frac{1}{2}(\partial_{x}\rho)^{2}+\frac{1}{2}\rho^{2}(\partial_{x}\theta)^{2}=\tilde{V}(\rho,\theta),
\end{equation}
where $\tilde{V}(\rho,\theta)\equiv V(\phi,\bar{\phi})$. In the case of $U(1)$ invariant potential, the equation of motion for $\theta$ is
\begin{equation}\label{2.28}
2\rho\partial_{x}\rho\partial_{x}\theta+\rho^{2}\partial_{x}^{2}\theta=0.
\end{equation}
Since for the static configuration, the $U(1)$ current $j_{x}=\rho^{2}\partial_{x}\theta$ is a constant which follows from (\ref{2.28}), the expression of $\frac{d\theta}{dx}$ depends only on $\rho^{2}$ inversely
\begin{equation}\label{2.29}
\frac{d\theta}{dx}=\frac{1}{\rho^{2}}j_{x}.
\end{equation}
As a result of the above equation, the equation (\ref{5.13}) becomes
\begin{equation}
\frac{d\rho}{dx}=\pm\sqrt{2\tilde{V}(\rho)-\frac{j_{x}^{2}}{\rho^{2}}},
\end{equation}
which can be solved analytically for a given $U(1)$ invariant potential. Although the above equation seems to suggests the existence of static soliton configurations with constant $U(1)$ current where the positive sign denotes kink solution and the negative sign denotes anti-kink solution, however, $j_{x}$ has to be zero which follows from the finiteness of energy of a static soliton configuration. This can also be explained using the fact that the roots of $\tilde{V}(\rho)$ and $2\tilde{V}(\rho)-\frac{j_{x}^{2}}{\rho^{2}}$ are not the same if $j_{x}\neq0$. This leads us to the trivial solution in which $\theta(x)$ is constant that implies the solutions of $\phi_{1}(x)$ and $\phi_{2}(x)$ are the same up to a multiplicative factor. Hence, the trivial solution of soliton configuration is the only solution for the $SO(2)$ or $U(1)$ invariant scalar field theory.

\section{MSTB model}\label{section 3}
\subsection{Introduction to the model}
MSTB model, shown in \cite{sarker1976solitary, montonen1976solitons, subbaswamy1981intriguing, subbaswamy1980instability, izquierdo2008generalized, izquierdo2019kink} can be described by adding a quadratic term in $\phi_{2}$ field in the action (\ref{eqn.0.1}) which breaks the $SO(2)$ invariance. The remaining symmetry of the theory is the discrete $\mathbb{Z}(2)$ symmetry. The corresponding action is
\begin{equation}\label{action-MSTB1}
\begin{split}
S[\phi_{1},\phi_{2}] & =\int d^{4}x\Big[\frac{1}{2}\sum_{a=1}^{2}\partial_{\mu}\phi_{a}\partial^{\mu}\phi_{a}-U(\{\phi_{a}\})\Big]\\
U(\{\phi_{a}\}) & =\frac{\lambda}{4}(\phi_{1}^{2}+\phi_{2}^{2}-\eta^{2})^{2}+\frac{1}{2}M^{2}\phi_{2}^{2},
\end{split}
\end{equation}
and the Euler-Lagrange equations of motion become the following
\begin{equation}
\begin{split}
\partial_{\mu}\partial^{\mu}\phi_{1} & =-\lambda\phi_{1}(\phi_{1}^{2}+\phi_{2}^{2}-\eta^{2})\\
\partial_{\mu}\partial^{\mu}\phi_{2} & =-\lambda\phi_{2}(\phi_{1}^{2}+\phi_{2}^{2}-\eta^{2}+\frac{M^{2}}{\lambda}).
\end{split}
\end{equation}
The above potential in the MSTB model can also be expressed in the form of (\ref{0.0}), shown in \cite{izquierdo2019kink} using the elliptic coordinates. However, we follow a completely different approach. The potential in the MSTB model is shown in Fig.2.
\begin{figure}
\begin{center}
\includegraphics[height=7cm,width=9cm]{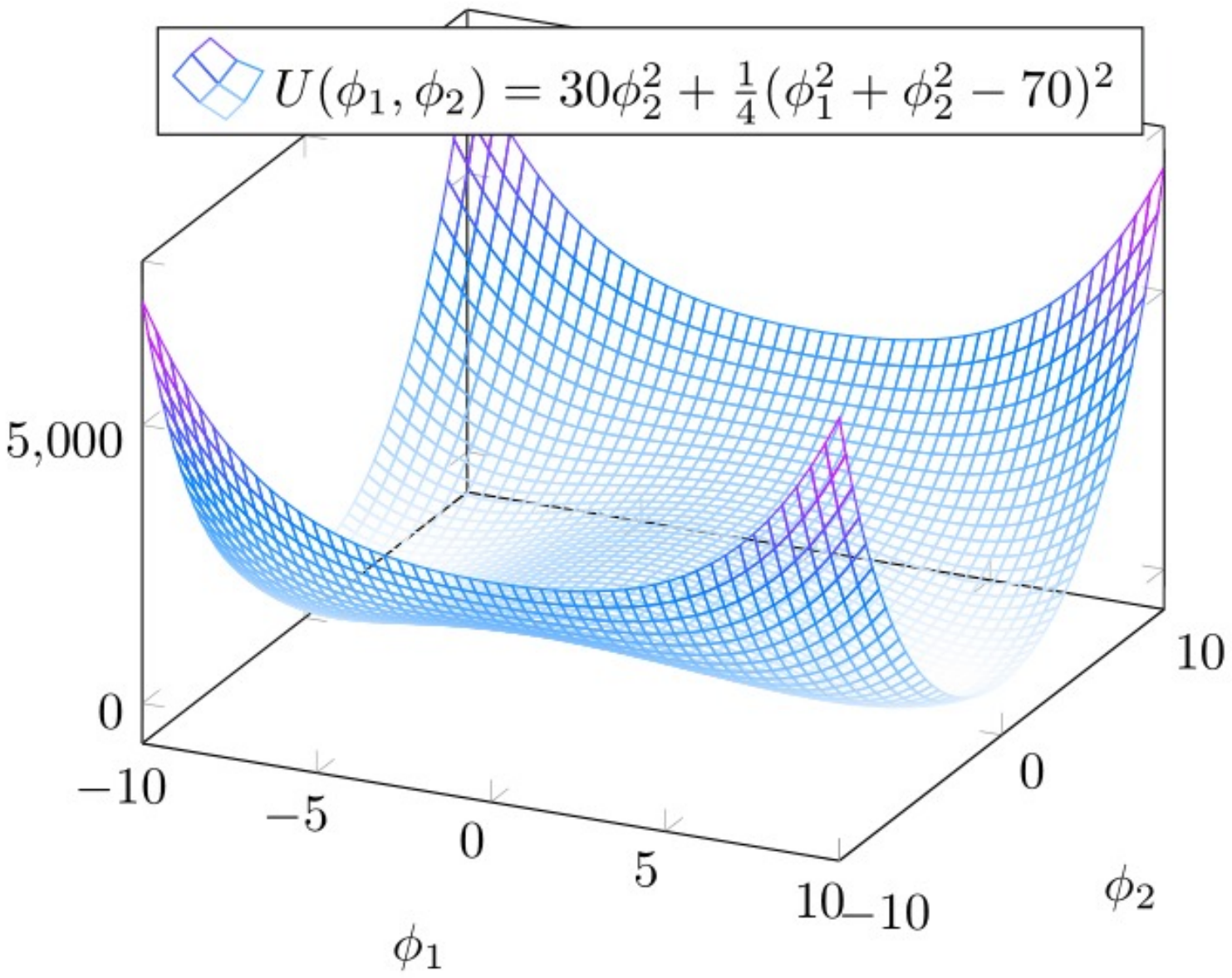}
\end{center}
\caption{Example of a potential in the MSTB model}
\end{figure}

\subsection{Analysis using the polar parametrization}
Using the polar parametrization, we obtain the following Lagrangian density
\begin{equation}\label{3.14}
\mathcal{L}=\frac{1}{2}(\partial\rho)^{2}+\frac{1}{2}\rho^{2}(\partial\theta)^{2}-\frac{\lambda}{4}(\rho^{2}-\eta^{2})^{2}-\frac{M^{2}}{2}\rho^{2}\sin^{2}\theta,
\end{equation}
for the MSTB model. The expression for energy corresponding to a static soliton configuration is given by
\begin{equation}\label{MSTB energy}
E=\int_{-\infty}^{\infty}dx\Big[\frac{1}{2}(\partial_{x}\rho)^{2}+\frac{1}{2}(\rho\partial_{x}\theta)^{2}+\frac{\lambda}{4}(\rho^{2}-\eta^{2})^{2}+\frac{M^{2}}{2}\rho^{2}\sin^{2}\theta\Big],
\end{equation}
and a finite-energy configuration must hold the following conditions
\begin{equation}
\begin{split}
E<\infty & \implies\partial_{x}\rho|_{x=\pm\infty}=0, \ \partial_{x}\theta|_{x=\pm\infty}=0\\
\rho(x=\pm\infty)=\eta, & \ \theta(x=\pm\infty)=m_{\pm}\pi,
\end{split}
\end{equation}
where $m_{\pm}\in\mathbb{Z}$ and $m=m_{+}-m_{-}$ is a topological number. The field equations in static limit become 
\begin{equation}\label{3.17}
\begin{split}
\partial_{x}^{2}\rho & =\rho(\partial_{x}\theta)^{2}+\lambda\rho(\rho^{2}-\eta^{2})+M^{2}\rho\sin^{2}\theta\\
\rho^{2}\partial_{x}^{2}\theta & +2\rho\partial_{x}\rho\partial_{x}\theta=\frac{M^{2}}{2}\rho^{2}\sin2\theta\\
\implies\partial_{x}\Big[\frac{1}{2}(\partial_{x}\rho)^{2} & +\frac{1}{2}\rho^{2}(\partial_{x}\theta)^{2}-\frac{\lambda}{4}(\rho^{2}-\eta^{2})^{2}-\frac{M^{2}}{2}\rho^{2}\sin^{2}\theta\Big]=0.
\end{split}
\end{equation}
The asymptotic conditions for the vacuum manifold demands
\begin{equation}\label{3.18}
\frac{1}{2}(\partial_{x}\rho)^{2}+\frac{1}{2}\rho^{2}(\partial_{x}\theta)^{2}=\frac{\lambda}{4}(\rho^{2}-\eta^{2})^{2}+\frac{M^{2}}{2}\rho^{2}\sin^{2}\theta.
\end{equation}

It is important to note that if $\rho(x)>\eta$ for all values of $x$, then
\begin{equation}
\partial_{x}^{2}\rho=\rho(\partial_{x}\theta)^{2}+\lambda\rho(\rho^{2}-\eta^{2})+M^{2}\rho\sin^{2}\theta>0.
\end{equation}
Since $\partial_{x}\rho|_{x=\pm\infty}=0$, $\partial_{x}\rho$ either increases first and then decreases or follow the reverse order. However, if $\rho(x)<\eta$ then
\begin{equation}
\partial_{x}^{2}\rho=\rho(\partial_{x}\theta)^{2}+M^{2}\rho\sin^{2}
\theta-\lambda\rho|(\rho^{2}-\eta^{2})|>0 \ \text{or} \ <0,
\end{equation} 
which means that depending on the value of $x$, $\partial_{x}\rho$ can either be increasing or decreasing, which shows $\rho(x)\leq\eta$ must be a condition that a soliton configuration must follow.

\subsection{Solitons in the MSTB model}
In order to find out the non-trivial finite-energy solutions of the MSTB model, a system of coupled first-order ODEs is established here using the non-conserved $U(1)$-current. Since this current is not conserved as the action of the MSTB model is not $U(1)$ invariant, the spatial variation of the current would be non-trivial. We also explicitly show the well-known analytic solutions of the MSTB model can be obtained from this system of first-order ODEs, shown below. 

The $U(1)$ current $j_{x}$ is defined in terms of the spatial derivative of $\theta$ in the following way
\begin{equation}\label{3.29}
j_{x}=\rho^{2}\frac{d\theta}{dx}.
\end{equation}
On the other hand, the equation (\ref{3.18}) can be expressed as
\begin{equation}\label{3.28}
\begin{split}
\Big[\left(\frac{d\rho}{dx}\right)^{2} & -\frac{\lambda}{2}(\rho^{2}-\eta^{2})^{2}\Big]=\Big[-\frac{j_{x}^{2}}{\rho^{2}}+M^{2}\rho^{2}\sin^{2}\theta\Big]\\
\implies\frac{d\rho}{dx} & =\pm\sqrt{\frac{\lambda}{2}(\rho^{2}-\eta^{2})^{2}-\frac{j_{x}^{2}}{\rho^{2}}+M^{2}\rho^{2}\sin^{2}\theta},
\end{split}
\end{equation}
where \textit{l.h.s} depends only on $\rho$, and \textit{r.h.s} depends on $\theta$ and current $j_{x}$. Further, in the \textit{r.h.s}, the positive sign denotes kink soliton solution and the negative sign denotes anti-kink soliton solution. It is important to note from the above equation that in order to satisfy the asymptotic conditions for finite-energy configurations, $\theta(x\rightarrow\pm\infty)\equiv\theta_{\pm}=m_{\pm}\pi$ must hold where $m_{\pm}\in\mathbb{Z}$ and $m\equiv m_{+}-m_{-}$ is known as the winding number of solitons. In a similar manner, the spatial derivative of $j_{x}$ can be expressed as follows
\begin{equation}\label{3.30}
\frac{dj_{x}}{dx}=\frac{M^{2}\rho^{2}}{2}\sin2\theta,
\end{equation}
which follows from the second equation in (\ref{3.17}). Hence, we obtain three first-order coupled ODEs (\ref{3.29}), (\ref{3.28}) and (\ref{3.30}) with the boundary conditions $\rho(x\rightarrow\pm\infty)=\pm\eta$, $j_{x}(x\rightarrow\pm\infty)=0$.

In order to find the kink and anti-kink configurations numerically in any classical field theory, the field equations must be solved within the right domain of initial conditions as the implementation of asymptotic boundary conditions are often difficult. In order to solve the first-order ODEs in (\ref{3.29}, \ref{3.28}, \ref{3.30}), we need certain initial conditions such that the asymptotic boundary conditions must be satisfied which is very hard to find in the MSTB model. However, the  following analytic solutions for $\lambda=2$, $\eta=1$ and $M\in(-1,1)$, shown in \cite{izquierdo2008generalized} 
\begin{equation}
\phi_{1}(x)=\tanh(Mx), \ \phi_{2}(x)=\pm\bar{M}\sech(Mx)
\end{equation}
are indeed the solutions of the above system of first-order differential equations where $\bar{M}=\sqrt{1-M^{2}}$. This can be checked easily. This soliton configuration is shown in Fig.3. Hence, the non-analytical solutions of the MSTB model can also be obtained by solving the above first-order differential equations with well-defined boundary conditions such that the following expression
\begin{equation}
\frac{\lambda}{2}(\rho^{2}-\eta^{2})^{2}-\frac{j_{x}^{2}}{\rho^{2}}+M^{2}\rho^{2}\sin^{2}\theta,
\end{equation}
remains positive-definite and becomes zero as $x\rightarrow\pm\infty$.
\begin{figure}
\includegraphics[height=4.5cm,width=5.9cm]{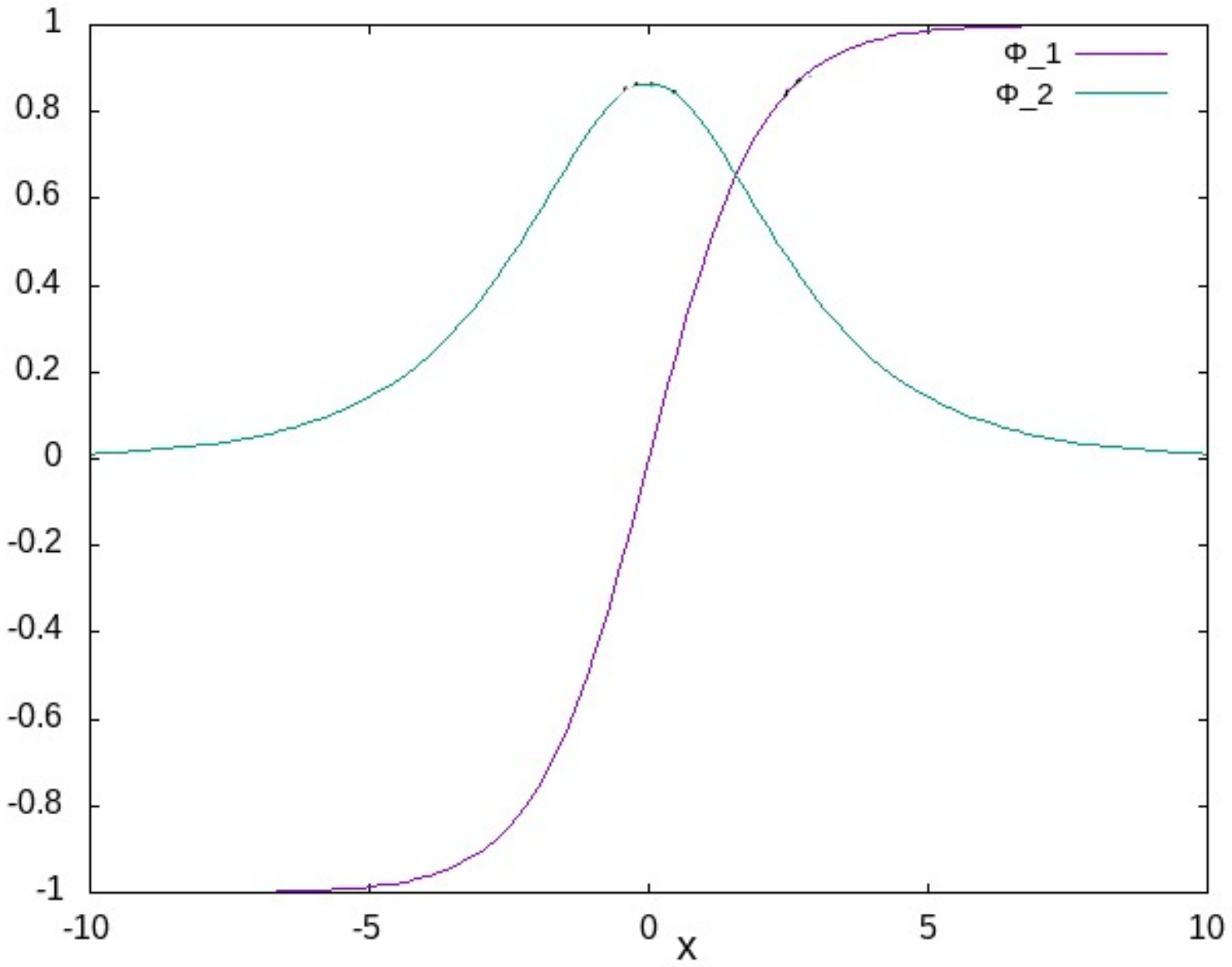}
\includegraphics[height=4.5cm,width=5.9cm]{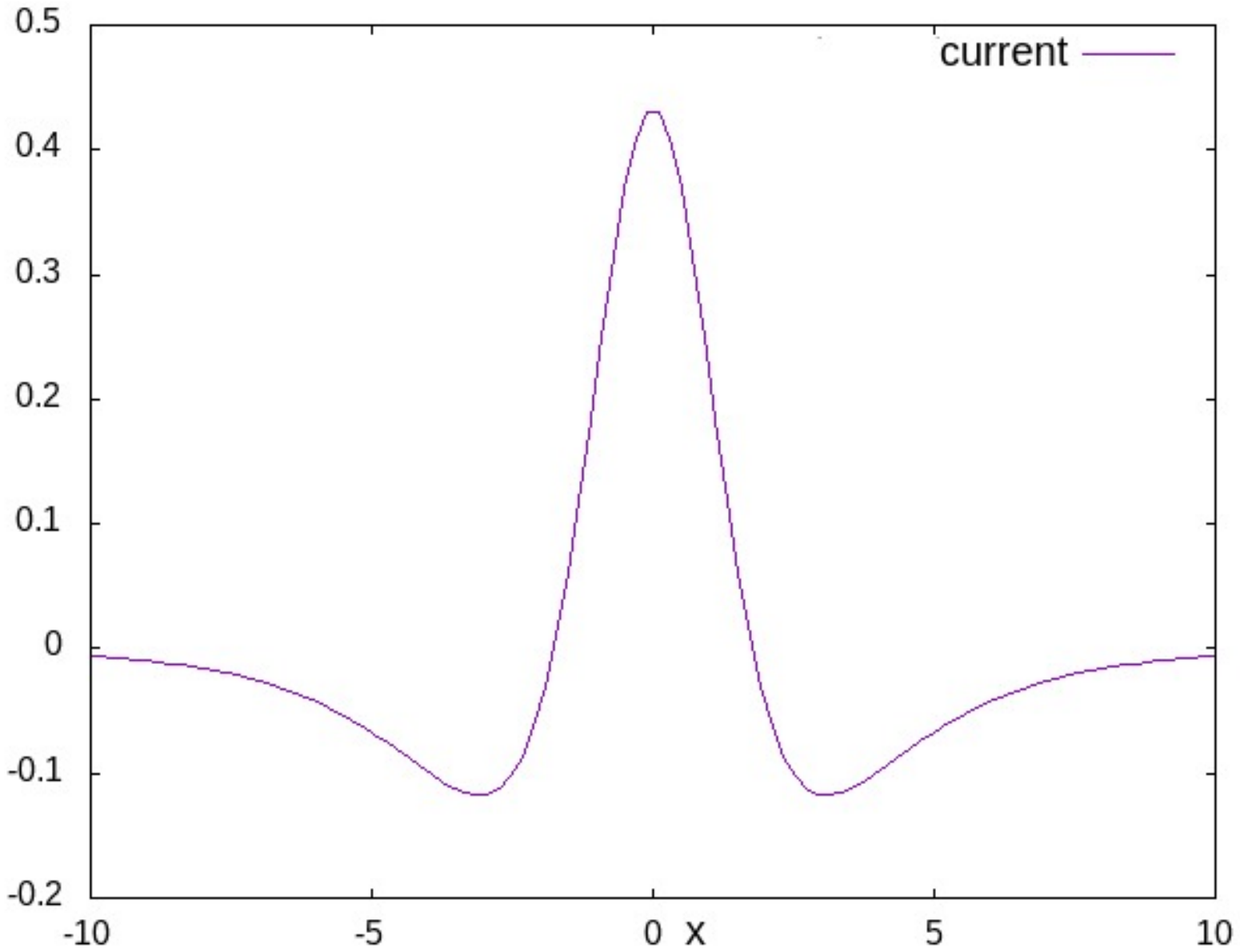}
\caption{Plot of an analytic finite-energy solution $\phi_{1}(x)=\tanh(Mx), \ \phi_{2}(x)=\pm\bar{M}\sech(Mx)$ in the MSTB model with the current.}
\end{figure}

An approximate finite-energy solution of the Euler-Lagrange equations can be obtained using the WKB approximation. Within this approximation $\left(\frac{d\theta}{dx}\right)^{2}\ll 1$, the Euler-Lagrange equations (\ref{3.17}) become 
\begin{equation}\label{WKB}
\begin{split}
\frac{d^{2}\rho}{dx^{2}} & =\rho\left(\frac{d\theta}{dx}\right)^{2}+\lambda\rho(\rho^{2}-\eta^{2})+M^{2}\rho\sin^{2}\theta\\ 
\frac{d^{2}\theta}{dx^{2}} & =\frac{M^{2}}{2}\sin2\theta\implies\frac{d\theta}{dx}=\pm M\sin\theta.
\end{split}
\end{equation}
The solution of the second ODE with positive sign is 
\begin{equation}
\theta(x)=\arcsec\left(\frac{\mathcal{M}_{0}e^{-2M(x-x_{0})}+1}{\mathcal{M}_{0}e^{-2M(x-x_{0})}-1}\right), \ \mathcal{M}_{0}=\left(\frac{1+\cos\theta_{0}}{1-\cos\theta_{0}}\right),
\end{equation}
where $\theta(x_{0})=\theta_{0}$. Then, the first ODE in (\ref{WKB}) becomes
\begin{equation}
\begin{split}
\frac{d^{2}\rho}{dx^{2}} & =\lambda\rho(\rho^{2}-\eta^{2})+2M^{2}\rho\sin^{2}\theta\\
 & =\lambda\rho(\rho^{2}-\eta^{2})+2M^{2}\rho\sech^{2}(M(x-x_{0}-\psi)),
\end{split}
\end{equation}
where $e^{-M\psi}=\frac{1}{\sqrt{\mathcal{M}_{0}}}$. Since $M^{2}\rho\sin^{2}\theta=\rho\left(\frac{d\theta}{dx}\right)^{2}\ll\rho$, and in order to obtain a finite-energy configuration, the equation (\ref{3.18}) must hold, we can make the following approximation
\begin{equation}
\frac{d^{2}\rho}{dx^{2}}\approx\lambda\rho(\rho^{2}-\eta^{2})\implies\rho(x)=\eta\left(\frac{\mathcal{P}_{0}e^{-\sqrt{2\lambda}\eta(x-x_{0})}+1}{\mathcal{P}_{0}e^{-\sqrt{2\lambda}\eta(x-x_{0})}-1}\right), \ \mathcal{P}_{0}=\frac{\rho_{0}+\eta}{\rho_{0}-\eta},
\end{equation}
where $\rho(x_{0})=\rho_{0}$. This solution is shown in Fig.4 with the spatial configuration of $U(1)$ current and the energy density. These solutions are non-topological since their topological charges are zero. The length scales involved in the WKB solutions of $\rho(x)$ and $\theta(x)$ are $\frac{1}{\sqrt{2\lambda}\eta}$ and $\frac{1}{2M}$ respectively. We define the following observable
\begin{equation}
\mathcal{Q}(x)=\int_{-\infty}^{x}j_{x}(y)dy,
\end{equation}
which is also plotted with the spatial derivative of the current in Fig.4.

\begin{figure}
\begin{center}
\includegraphics[height=4.7cm,width=5.9cm]{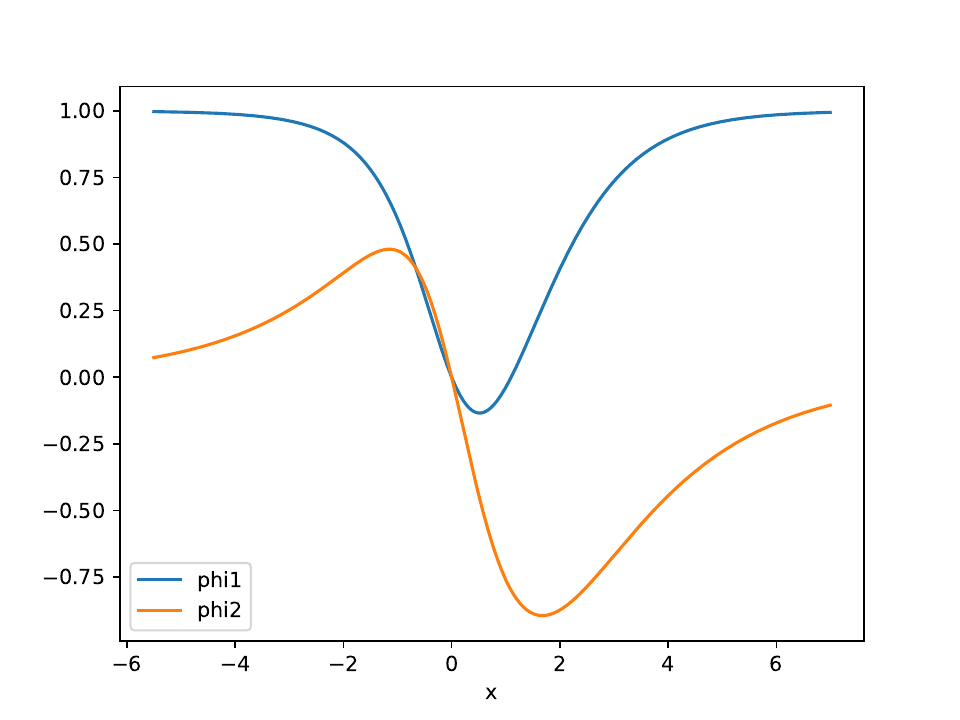}
\includegraphics[height=4.7cm,width=5.9cm]{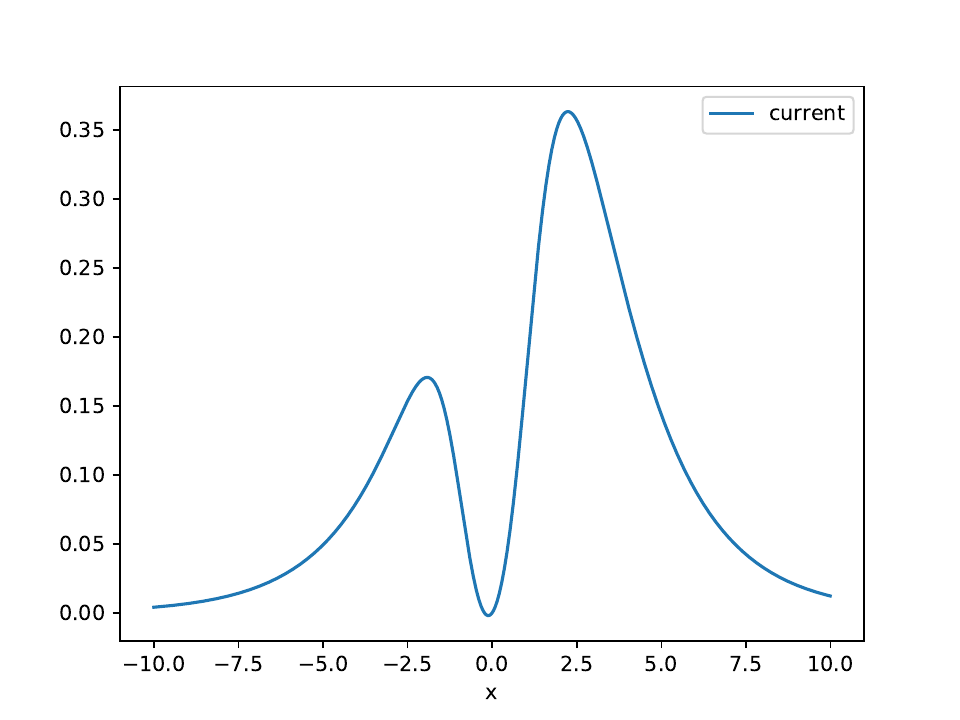}\\
\includegraphics[height=4.7cm,width=5.9cm]{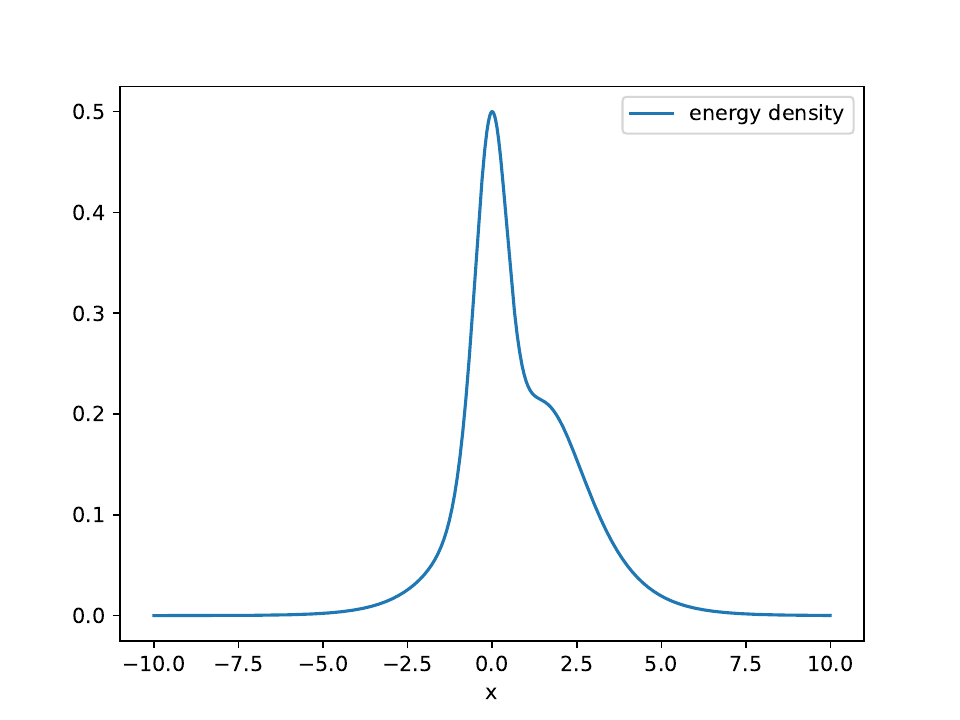}
\includegraphics[height=4.7cm,width=5.9cm]{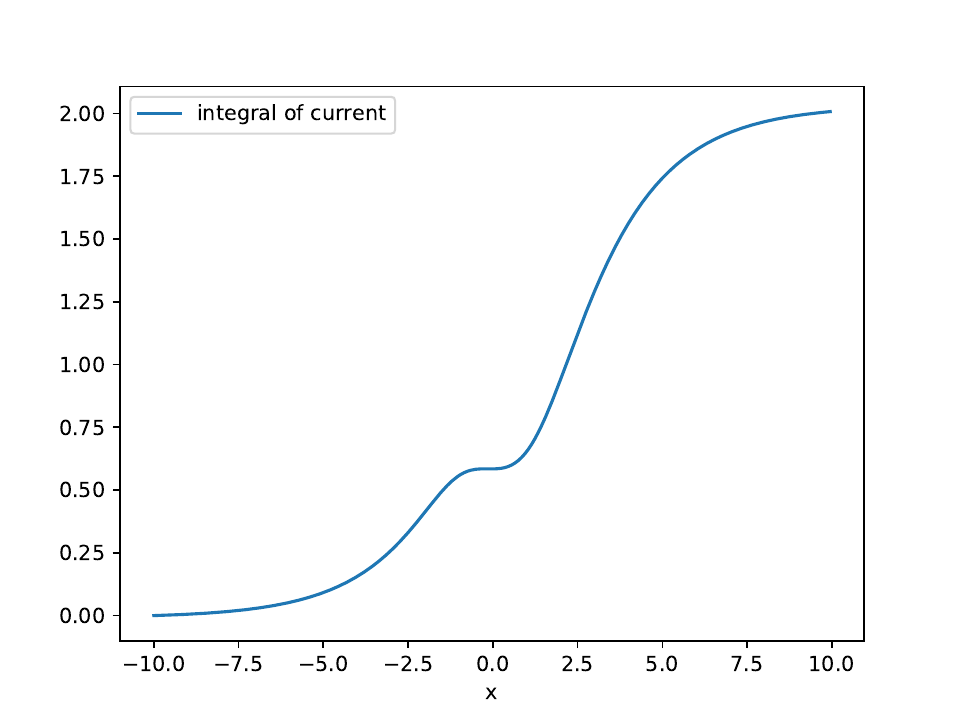}\\
\includegraphics[height=4.7cm,width=9cm]{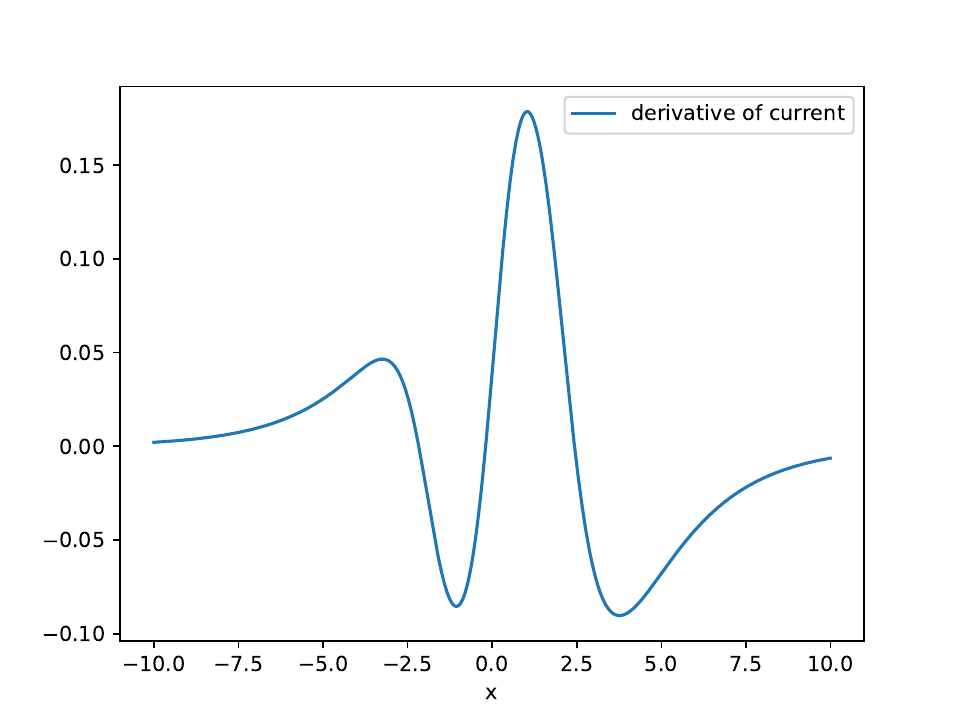}
\end{center}
\caption{Soliton configuration in MSTB model with $\lambda=2, \ \eta=1, \ m=\frac{1}{2}$ in WKB approximation.}
\end{figure}

From the system of first-order ODEs (\ref{3.29}, \ref{3.28}, \ref{3.30}), the asymptotic behavior of the field variables can also be obtained, shown below. The equation (\ref{3.28}) can also be expressed in the following ways
\begin{equation}\label{complex-analysis}
\begin{split}
\left(\frac{d\rho}{dx}-i\frac{j_{x}}{\rho}\right)\left(\frac{d\rho}{dx}+i\frac{j_{x}}{\rho}\right) & =\left(\sqrt{\frac{\lambda}{2}}(\rho^{2}-\eta^{2})-iM\rho\sin\theta\right)\left(\sqrt{\frac{\lambda}{2}}(\rho^{2}-\eta^{2})+iM\rho\sin\theta\right)\\
\left(\frac{d\rho}{dx}-i\frac{j_{x}}{\rho}\right)\left(\frac{d\rho}{dx}+i\frac{j_{x}}{\rho}\right) & =\left(-\sqrt{\frac{\lambda}{2}}(\rho^{2}-\eta^{2})-iM\rho\sin\theta\right)\left(-\sqrt{\frac{\lambda}{2}}(\rho^{2}-\eta^{2})+iM\rho\sin\theta\right).
\end{split}
\end{equation}
The positive and negative signs in \textit{r.h.s} denote kink and anti-kink solitons respectively. For the time being, we consider the first equation in (\ref{complex-analysis}) to find the asymptotic behavior of soliton configuration in the MSTB model. The general solution of the first equation is 
\begin{equation}
\frac{d\rho}{dx}+i\frac{j_{x}}{\rho}=\left(\sqrt{\frac{\lambda}{2}}(\rho^{2}-\eta^{2})\pm iM\rho\sin\theta\right)e^{i\Phi(x)},
\end{equation}
where $e^{i\Phi(x\rightarrow\pm\infty)}=1$ which we show now. If $e^{i\Phi(x\rightarrow\pm\infty)}=1$ holds, then the above equation gives the following conditions
\begin{equation}
\frac{d\rho}{dx}=\sqrt{\frac{\lambda}{2}}(\rho^{2}-\eta^{2}), \ \frac{d\theta}{dx}=M\sin\theta,
\end{equation}
in $x\rightarrow\pm\infty$ limits and it can also be checked easily that the equation (\ref{3.30}) also holds in these limits. The solutions of the  above two equations in $x\rightarrow\pm\infty$ with the boundary conditions mentioned earlier are given by 
\begin{equation}
\rho(x)=\eta\left(\frac{\bar{\mathcal{P}}_{0}e^{-\sqrt{2\lambda}\eta(x-x_{0})}+1}{\bar{\mathcal{P}}_{0}e^{-\sqrt{2\lambda}\eta(x-x_{0})}-1}\right), \ \bar{\mathcal{P}}_{0}=\frac{\rho_{0}+\eta}{\rho_{0}-\eta},
\end{equation}
where $\rho(x_{0})=\rho_{0}$, and
\begin{equation}
\theta(x)=\arcsec\left(\frac{\bar{\mathcal{M}}_{0}e^{-2M(x-x_{0})}+1}{\bar{\mathcal{M}}_{0}e^{-2M(x-x_{0})}-1}\right), \ \bar{\mathcal{M}}_{0}=\left(\frac{1+\cos\theta_{0}}{1-\cos\theta_{0}}\right),
\end{equation}
where $\theta(x_{0})=\theta_{0}$ and $|x_{0}|\gg0$. Hence, the expression for current in $x\rightarrow\pm\infty$ with the above boundary conditions is given by
\begin{equation}
j_{x}(x)=M\rho^{2}(x)\sin\theta(x).
\end{equation}
However, from (\ref{complex-analysis}), we obtain the following two relations
\begin{equation}\label{complex2}
\frac{d\rho}{dx}=\sqrt{\frac{\lambda}{2}}(\rho^{2}-\eta^{2})\cos\Phi-M\rho\sin\theta\sin\Phi, \ \frac{d\theta}{dx}=\frac{1}{\rho}\sqrt{\frac{\lambda}{2}}(\rho^{2}-\eta^{2})\sin\Phi+M\sin\theta\cos\Phi
\end{equation}
that must be solved in order to find out the soliton configurations. Since soliton configurations are the solutions of Euler-Lagrange equations given in (\ref{3.17}), we obtain the following constraints
\begin{equation}
\begin{split}
\lambda\rho & (\rho^{2}-\eta^{2})+M^{2}\rho\sin\theta=\frac{d^{2}\rho}{dx^{2}}-\rho\left(\frac{d\theta}{dx}\right)^{2}\\
 & =\frac{d\rho}{dx}[\sqrt{2\lambda}\rho\cos\Phi-M\sin\theta\sin\Phi]-M\rho\cos\theta\sin\Phi\frac{d\theta}{dx}-\rho\frac{d\theta}{dx}\frac{d\Phi}{dx}-\rho\left(\frac{d\theta}{dx}\right)^{2},
\end{split}
\end{equation}
and 
\begin{equation}
\begin{split}
\frac{M^{2}\rho^{2}\sin2\theta}{2} & =\frac{d}{dx}\left(\rho^{2}\frac{d\theta}{dx}\right)\\
 & =\frac{d\rho}{dx}\Big[\sqrt{\frac{\lambda}{2}}(\rho^{2}-\eta^{2})\sin\Phi+\sqrt{2\lambda}\rho^{2}\sin\Phi+2M\rho\sin\theta\cos\Phi\Big]\\
 & +M\rho^{2}\cos\theta\cos\Phi\frac{d\theta}{dx}+\rho\frac{d\rho}{dx}\frac{d\Phi}{dx}.
\end{split}
\end{equation}
The above two relations are compatible with each other if and only if the following condition holds
\begin{equation}
\begin{split}
\frac{d\rho}{d\theta}[\sqrt{2\lambda} & \rho\cos\Phi-M\sin\theta\sin\Phi]-M\rho\cos\theta\sin\Phi-\frac{1}{\frac{d\theta}{dx}}[\lambda\rho(\rho^{2}-\eta^{2})+M^{2}\rho\sin\theta]-\rho\frac{d\theta}{dx}\\
=\rho\frac{d\Phi}{dx} & =\frac{1}{\frac{d\rho}{dx}}\frac{M^{2}\rho^{2}\sin2\theta}{2}-\Big[\sqrt{\frac{\lambda}{2}}(\rho^{2}-\eta^{2})\sin\Phi+\sqrt{2\lambda}\rho^{2}\sin\Phi+2M\rho\sin\theta\cos\Phi\Big]\\
 & -M\rho^{2}\cos\theta\cos\Phi\frac{d\theta}{d\rho},
\end{split}
\end{equation}
which acts as a constraint. Since $\frac{d\rho}{dx}, \ \frac{d\theta}{dx}$, and $\frac{d\rho}{d\theta}$ are known from the equation (\ref{complex2}), hence, the above relation can be solved in principle in terms of $\Phi$ as a function of $\rho$ and $\theta$. Then substituting the function $\cos\Phi[\rho,\theta]$ and $\sin\Phi[\rho,\theta]$ obtained from the above constraint in (\ref{complex2}), we obtain two first-order differential equations with asymptotic boundary conditions which can be solved numerically as a boundary value problem. Therefore, using this approach, we reduced the number of ODEs governing the spatial behavior of a soliton configuration in this model.

Using the system of first-order ODEs (\ref{3.29}, \ref{3.28}, \ref{3.30}), the expressions of current $j_{x}$, and $\theta$ in terms of the field variable $\rho$ can be obtained by solving the following first-order differential equations 
\begin{equation}\label{-3.31}
\begin{split}
\frac{dj_{x}}{d\rho} & =\pm\frac{M^{2}\rho^{2}}{2\sqrt{\frac{\lambda}{2}(\rho^{2}-\eta^{2})^{2}-\frac{j_{x}^{2}}{\rho^{2}}+M^{2}\rho^{2}\sin^{2}\theta}}\\
\frac{d\theta}{d\rho} & =\pm\frac{j_{x}}{\rho^{2}\sqrt{\frac{\lambda}{2}(\rho^{2}-\eta^{2})^{2}-\frac{j_{x}^{2}}{\rho^{2}}+M^{2}\rho^{2}\sin^{2}\theta}},
\end{split}
\end{equation}
for kink and anti-kink configurations. Since the expressions of $\theta$, and current $j_{x}$ can be obtained in terms of $\rho$ by solving the above equations, the knowledge about the dynamics of $\rho$ field is sufficient to predict the dynamics of other field variables in the MSTB model. The dynamics of fundamental field variables contain all the necessary information about the observables in a classical field theory. However, that is not true for its corresponding quantum theory since aside from the field operators, the description of a quantum state in the Hilbert space is important in quantum field theories for describing the properties of a system. These properties are mainly encoded in the correlation functions of fundamental field operators in that theory. Since the excitations around a coherent state describing a classical soliton configuration (see \cite{mandal2019low, dvali2015towards}) are encoded in the dynamics of $\rho$ field operators, all the correlations or any other observable can be obtained from an effective field theory, described by the following action
\begin{equation}
S_{\rho}=\int d^{2}x\Big[\frac{1}{2}\partial_{\mu}\rho\partial^{\mu}\rho-\mathcal{V}(\rho)\Big], \ \mathcal{V}(\rho)=\frac{\lambda}{4}(\rho^{2}-\eta^{2})^{2}+\frac{1}{2}M^{2}\rho^{2}\sin^{2}\theta[\rho]-\frac{j_{x}^{2}[\rho]}{\rho^{2}}.
\end{equation}
The expression for effective potential $\mathcal{V}(\rho)$ in the above action shows explicitly that the $U(1)$ current gives rise to an attraction term in $\mathcal{V}(\rho)$. Further, the excitations of $\rho$ field variable around the kink and anti-kink configurations determine the excitation of $\theta$ field variable which follows from the equations (\ref{-3.31}).

Moreover, if the vacuum manifold (the internal space of fields for which the potential becomes vanishing) of a given potential is known exactly, then the equations (\ref{-3.31}) can be used to determine $\theta$ and $j_{x}$ exactly in terms of $\rho$ connecting two points in the vacuum manifold asymptotically. Then the $\theta$ and $j_{x}$ as a function of $\rho$ can be used further in order to know spatial dependence of $\rho, \ \theta, \ j_{x}$. For this reason, this approach would be useful in finding the soliton configurations in classical field theories.

\section{Extension of the approach beyond the MSTB model}\label{section4}
The approach mentioned in the previous section can be used in finding the exact soliton configurations in different classes of classical field theories that are an extension of the MSTB model. The action of one such class of theories are of the following form
\begin{equation}
S=\int d^{2}x\Big[\frac{1}{2}\partial_{\mu}\rho\partial^{\mu}\rho+\frac{1}{2}\rho^{2}\partial_{\mu}\theta\partial^{\mu}\theta-\frac{\lambda}{2}\mathcal{U}^{2}(\rho)-\frac{M^{2}\rho^{2}}{2}\mathcal{K}^{2}(\theta)\Big],
\end{equation} 
where $U(\rho), \mathcal{K}(\theta)$ are two arbitrary functions that have non-trivial roots. The function $\mathcal{K}(\theta)$ can also be expressed in the canonical form as
\begin{equation}
\bar{\mathcal{K}}\left(\frac{\phi_{2}}{\phi_{1}}\right)=\mathcal{K}\left(\arctan\left(\frac{\phi_{2}}{\phi_{1}}\right)\right).
\end{equation} 

The Euler-Lagrange equations for static configurations corresponding to this class of theories are given by
\begin{equation}
\frac{d^{2}\rho}{dx^{2}}=\rho\left(\frac{d\theta}{dx}\right)^{2}+\lambda\mathcal{U}(\rho)\frac{d\mathcal{U}(\rho)}{d\rho}+M^{2}\rho\mathcal{K}^{2}(\theta), \ \frac{d}{dx}\left(\rho^{2}\frac{d\theta}{dx}\right)=M^{2}\rho^{2}\mathcal{K}(\theta)\frac{d\mathcal{K}(\theta)}{d\theta}.
\end{equation}
The asymptotic boundary conditions for the finite-energy configurations demand the following condition
\begin{equation}\label{condition class}
\left(\frac{d\rho}{dx}\right)^{2}+\rho^{2}\left(\frac{d\theta}{dx}\right)^{2}=\lambda\mathcal{U}^{2}(\rho)+M^{2}\rho^{2}\mathcal{K}^{2}(\theta).
\end{equation}
The expression for energy corresponding to a static finite-energy configuration is given by
\begin{equation}
E=\int_{-\infty}^{\infty}\Bigg[\frac{1}{2}\left(\frac{d\rho}{dx}\right)^{2}+\frac{\rho^{2}}{2}\left(\frac{d\theta}{dx}\right)^{2}+\frac{\lambda}{2}\mathcal{U}^{2}(\rho)+\frac{M^{2}\rho^{2}}{2}\mathcal{K}^{2}(\theta)\Bigg]dx.
\end{equation}
Within the WKB approximation, the following ODEs must be solved in order to obtain a soliton configuration
\begin{equation}
\frac{d\rho}{dx}=\pm\sqrt{\lambda\mathcal{U}(\rho)}, \ \frac{d\theta}{dx}=M\mathcal{K}(\theta).
\end{equation}
Like the MSTB model, here the equation (\ref{condition class}) can be expressed in the following two ways
\begin{equation}
\begin{split}
\left(\frac{d\rho}{dx}-i\rho\frac{d\theta}{dx}\right)\left(\frac{d\rho}{dx}+i\rho\frac{d\theta}{dx}\right) & =\left(\sqrt{\frac{\lambda}{2}}\mathcal{U}(\rho)-iM\rho\mathcal{K}(\theta)\right)\left(\sqrt{\frac{\lambda}{2}}\mathcal{U}(\rho)+iM\rho\mathcal{K}(\theta)\right)\\
\left(\frac{d\rho}{dx}-i\rho\frac{d\theta}{dx}\right)\left(\frac{d\rho}{dx}+i\rho\frac{d\theta}{dx}\right) & =\left(-\sqrt{\frac{\lambda}{2}}\mathcal{U}(\rho)-iM\rho\mathcal{K}(\theta)\right)\left(-\sqrt{\frac{\lambda}{2}}\mathcal{U}(\rho)+iM\rho\mathcal{K}(\theta)\right).
\end{split}
\end{equation}
Hence, in the asymptotic limit $x\rightarrow\pm\infty$, soliton configurations behave as the solution of the following ODEs
\begin{equation}
\frac{d\rho}{dx}=\pm\sqrt{\frac{\lambda}{2}}\mathcal{U}(\rho), \ \frac{d\theta}{dx}=\pm M\mathcal{K}(\theta).
\end{equation}
The solutions of the above equations can be obtained exactly given the form of the function $\mathcal{K}(\theta)$. In order to find a soliton configuration in the domain $x\in(-\infty,\infty)$, the following ODEs must be solved
\begin{equation}\label{general ODEs}
\begin{split}
\frac{d\rho}{dx} & =\sqrt{\frac{\lambda}{2}}\mathcal{U}(\rho)\cos\Phi-M\rho\mathcal{K}(\theta)\sin\Phi\\
\frac{d\theta}{dx} & =\frac{1}{\rho}\sqrt{\frac{\lambda}{2}}\mathcal{U}(\rho)\sin\Phi+M\mathcal{K}(\theta)\cos\Phi,
\end{split}
\end{equation}
with the asymptotic boundary conditions mentioned earlier as a boundary value problem. The function $\Phi$ in terms of field variables $\rho, \ \theta$ can be obtained by solving the following relation
\begin{equation}
\begin{split}
 & \left(\sqrt{\frac{\lambda}{2}}\frac{d\mathcal{U}(\rho)}{d\rho}\cos\Phi-M\mathcal{K}(\theta)\sin\Phi\right)\frac{d\rho}{d\theta}-M\rho\frac{d\mathcal{K}(\theta)}{d\theta}\sin\Phi-\rho\frac{d\theta}{dx}-\lambda\mathcal{U}(\rho)\frac{d\mathcal{U}(\rho)}{d\rho}-M^{2}\rho\mathcal{K}^{2}(\theta)\\
= & \frac{M^{2}\rho^{2}\mathcal{K}(\theta)\frac{d\mathcal{K}(\theta)}{d\theta}}{\frac{d\rho}{dx}}-\left(\sqrt{\frac{\lambda}{2}}\mathcal{U}(\rho)\sin\Phi+\sqrt{\frac{\lambda}{2}}\rho\frac{d\mathcal{U}(\rho)}{d\rho}\sin\Phi+2M\rho\mathcal{K}(\theta)\cos\Phi\right)\\
 & -M\rho^{2}\frac{d\mathcal{K}(\theta)}{d\theta}\cos\Phi\frac{d\theta}{d\rho}.
\end{split}
\end{equation}
Therefore, a finite-energy solution connecting two fixed points asymptotically can be obtained by solving the first-order coupled differential equations (\ref{general ODEs}). Further, soliton configurations in the class of field theories described by the action (\ref{0.0}) can also be obtained by solving the following two coupled first-order ODEs
\begin{equation}
\begin{split}
\frac{d\phi_{1}}{dx} & =\cos\Phi\mathcal{W}_{\phi_{1}}-\sin\Phi\mathcal{W}_{\phi_{2}}\\
\frac{d\phi_{2}}{dx} & =\sin\Phi\mathcal{W}_{\phi_{1}}+\cos\Phi\mathcal{W}_{\phi_{2}},
\end{split}
\end{equation}
with the well-defined boundary conditions whereas the functional form of $\Phi(\phi_{1},\phi_{2})$ is obtained from the following relation
\begin{equation}
\begin{split}
\frac{\cos\Phi\mathcal{W}_{\phi_{1}}-\sin\Phi\mathcal{W}_{\phi_{2}}}{\sin\Phi\mathcal{W}_{\phi_{1}}+\cos\Phi\mathcal{W}_{\phi_{2}}} & \left(\cos\Phi\frac{\partial^{2}\mathcal{W}}{\partial\phi_{1}^{2}}-\sin\Phi\frac{\partial^{2}\mathcal{W}}{\partial\phi_{1}\partial\phi_{2}}\right)\\
+\left(\cos\Phi\frac{\partial^{2}\mathcal{W}}{\partial\phi_{1}\partial\phi_{2}}-\sin\Phi\frac{\partial^{2}\mathcal{W}}{\partial\phi_{2}^{2}}\right) & -\frac{\mathcal{W}_{\phi_{1}}\frac{\partial^{2}\mathcal{W}}{\partial\phi_{1}^{2}}+\mathcal{W}_{\phi_{2}}\frac{\partial^{2}\mathcal{W}}{\partial\phi_{1}\partial\phi_{2}}}{\sin\Phi\mathcal{W}_{\phi_{1}}+\cos\Phi\mathcal{W}_{\phi_{2}}}\\
 =\frac{\mathcal{W}_{\phi_{1}}\frac{\partial^{2}\mathcal{W}}{\partial\phi_{1}\partial\phi_{2}}+\mathcal{W}_{\phi_{2}}\frac{\partial^{2}\mathcal{W}}{\partial\phi_{2}^{2}}}{\cos\Phi\mathcal{W}_{\phi_{1}}-\sin\Phi\mathcal{W}_{\phi_{2}}} & -\left(\sin\Phi\frac{\partial^{2}\mathcal{W}}{\partial\phi_{1}^{2}}+\cos\Phi\frac{\partial^{2}\mathcal{W}}{\partial\phi_{1}\partial\phi_{2}}\right)\\
-\frac{\sin\Phi\mathcal{W}_{\phi_{1}}+\cos\Phi\mathcal{W}_{\phi_{2}}}{\cos\Phi\mathcal{W}_{\phi_{1}}-\sin\Phi\mathcal{W}_{\phi_{2}}} & \left(\sin\Phi\frac{\partial^{2}\mathcal{W}}{\partial\phi_{1}\partial\phi_{2}}+\cos\Phi\frac{\partial^{2}\mathcal{W}}{\partial\phi_{2}^{2}}\right).
\end{split}
\end{equation}
The expression for $U(1)$ current is given by 
\begin{equation}
\begin{split}
j_{x} & =\sin\Phi(\phi_{1}\mathcal{W}_{\phi_{1}}+\phi_{2}\mathcal{W}_{\phi_{2}})+\cos\Phi(\phi_{1}\mathcal{W}_{\phi_{2}}-\phi_{2}\mathcal{W}_{\phi_{1}})\\
 & =\Bigg[\sin\Phi\left(\phi_{1}\frac{\partial}{\partial\phi_{1}}+\phi_{2}\frac{\partial}{\partial\phi_{2}}\right)+\cos\Phi\left(\phi_{1}\frac{\partial}{\partial\phi_{2}}-\phi_{2}\frac{\partial}{\partial\phi_{1}}\right)\Bigg]\mathcal{W}.
\end{split}
\end{equation}
The operators in the parenthesis are defined over the internal space and similar to the operators $\vec{x}.\vec{\nabla}$ and $\vec{x}\times\vec{\nabla}$ in two-dimensional space.

\section{Discussion}
The existence of solitons leads to the occurrence of a broad class of physical phenomena \cite{bulanov2020electromagnetic, chai2020magnetic, mukherjee2019observation}. The study of coupled solitons in two interacting scalar field models \cite{saadatmand2014dynamics, morera2018dark, bazeia1995solitons} are mostly restricted within the numerical approaches since it is often difficult to find an analytic solution of a system of coupled first-order ODEs, governing the dynamics of the field variables. In this article, the existence of soliton-like finite-energy configurations is shown explicitly in a class of classical field theories containing two interacting fields through a system of coupled first-order ODEs governing the dynamics. Although the solution of soliton configuration in a $1+1$-dimension MSTB model is obtained from two coupled first-order ODEs in \cite{izquierdo2019kink} using the Euler coordinates, we follow a different approach in which $U(1)$ current is used explicitly. This approach can easily be generalized to different classes of field potentials in two interacting scalar field theories, shown in section \ref{section4}. Moreover, in this generalized approach, soliton configurations can be obtained by solving two coupled first-order ODEs rather than two second-order ODEs. In our approach, the existence of a non-trivial spatial configuration of current is also shown. Our approach is promising and helpful in order to understand solitons in the Ginzburg-Landau theories of different one-dimensional models \cite{efimkin2015moving, luther1977quantum, mikeska1981solitons, chiel2020symmetry}.

Further, through the MSTB model, it is shown explicitly that the classical dynamics can be determined entirely from one field variable. Since the dynamics of these field theories is entirely contained in one field variable shown explicitly, it is possible to write an effective action only in that field variable in order to study the excitations around the soliton configurations in its corresponding quantum version. As a consequence of this, all kinds of correlation functions in these field theories \textit{w.r.t} the coherent state corresponding to the classical soliton configuration can be obtained from the correlation functions in one field variable. Hence, our approach would also be helpful in understanding the dynamics of excitations around kink and anti-kink configurations in these coupled field theories. Furthermore, this approach can be used to find the finite-energy configurations in field theories containing potentials beyond the class of potentials of the form (\ref{0.0}).

\section{Acknowledgement}
SM wants to thank IISER Kolkata for supporting this work through a doctoral fellowship.

\bibliographystyle{unsrt}
\bibliography{draft}

\end{document}